\documentclass[a4paper,11pt]{article}
\pdfoutput=1
\usepackage{jcappub}
\usepackage{amssymb,amsmath,mathrsfs,enumerate}
\usepackage{graphicx,rotate,multicol}
\usepackage{float}

\usepackage{subfig}

\usepackage{slashed}
\usepackage{mathtools}
\usepackage{multirow}

\allowdisplaybreaks

\title{\boldmath PTOLEMY's test of generalized neutrino interactions: unveiling challenges and constraints}

\author[a]{Indra Kumar Banerjee,}
\author[a]{Ujjal Kumar Dey,}
\author[b]{Newton Nath,}
\author[a]{Saadat Salman Shariff}
\affiliation[a]{Department of Physical Sciences, Indian Institute of Science Education and Research Berhampur,\\Transit Campus, Government ITI, Berhampur 760010, Odisha, India}
\affiliation[b]{Istituto Nazionale di Fisica Nucleare - Sezione di Bari, \\ Via Orabona 4, 70126 Bari, Italy}

\emailAdd{indrab@iiserbpr.ac.in}
\emailAdd{ujjal@iiserbpr.ac.in}
\emailAdd{newton.nath@ba.infn.it}
\emailAdd{saadatsalman342@gmail.com}

\abstract{
Unanswered questions surrounding neutrinos have motivated investigations into physics beyond the standard model (SM) of particle physics. In particular, generalized neutrino interactions (GNI) provide a broader framework for studying these effects compared to the commonly studied non-standard neutrino interactions. These interactions are described by higher dimensional operators while maintaining the gauge symmetries of the SM.
Furthermore, the cosmic neutrino background, a predicted component of the SM and standard cosmology, has yet to be directly detected. To shed light on this elusive phenomenon, we conduct a comprehensive analysis of the relevant GNI, specifically focusing on their implications for the proposed cosmic neutrino detector PTOLEMY. We make an attempt to see the capabilities and the limitations of PTOLEMY in sensing GNI while remaining optimistic regarding PTOLEMY's experimental resolution. These interactions play a significant role in modifying the electron spectrum resulting from the capture of cosmic neutrinos on radioactive tritium.
This work also explores how the presence of these  interactions influences the differential electron spectrum, taking into account factors such as finite experimental resolution, the mass of the lightest neutrino eigenstate, the strength of the interactions, and the ordering of neutrino mass. 
}

\begin{document}
\maketitle
\flushbottom


\section{Introduction}
\label{sec:intro}
Among all hitherto known fundamental particles, neutrinos are the most recalcitrant ones to be tamed by the rules of the Standard Model (SM) of particle physics. The mere presence of minuscule mass, as indicated by the oscillation experiments, casts shadows on the completeness of the SM. This tiny but non-zero mass of neutrinos is arguably the most glaring drawback of the SM, and it asks for physics beyond the SM (BSM). Basically, in the SM, neutrinos are purely left-handed fermionic fields residing in the $SU(2)_L$ doublet and principles of gauge invariance prohibit mass terms for the neutrinos at the renormalizable level. BSM interactions of neutrinos can be conveniently parametrized in terms of a low-energy effective field theory of \textit{general neutrino interactions} (GNI), a subclass of which is popularly known as \textit{neutrino non-standard interactions} (NSI), see~\cite{Ohlsson:2012kf, Farzan:2017xzy, Proceedings:2019qno} for a review. In general, these GNIs can be generated in the form of dimension six~\cite{Buchmuller:1985jz, Bergmann:1998ft, Bergmann:1999pk, Wise:2014oea} and/or eight~\cite{Berezhiani:2001rs, Davidson:2003ha} effective operators.
%
A plethora of analyses  have been made in recent times to explore these interactions in neutrino oscillations \cite{Bischer:2018zcz, Khan:2019jvr}, and neutrino scattering \cite{Rodejohann:2017vup, Kosmas:2017tsq,AristizabalSierra:2018eqm,Chen:2021uuw,Li:2020lba,Flores:2021kzl,Nath:2022cur} experiments.  The latest global constraints of GNI parameters can be found in~\cite{Escrihuela:2021mud}.
While neutrinos are propelling a vibrant program in the particle physics sector, they play an equally crucial role in the frontiers of cosmology. The standard model of cosmology, i.e., the $\Lambda$CDM, predicts in the same vein as cosmic microwave background (CMB) the presence of a cosmic neutrino background (CNB)~\cite{Dicke:1965zz}. According to the $\Lambda$CDM, the neutrinos decoupled from other particles earlier ($\sim 1$ s after the Big Bang) than the CMB photons ($\sim 4\times 10^5$ yrs) and formed CNB. Detection of these neutrinos can reveal to us information about the earliest possible epoch of the Universe after the Big Bang. The contribution of the CNB neutrinos in the energy density of the Universe affects the light element abundances during nucleosynthesis and leaves its footprints in the CMB anisotropies, and structure formation. These observations form some kind of indirect signatures of CNB. However, due to the tiny masses, small interaction cross sections, and, low-temperature today of the CNB ($T_{\nu,0} \sim 1.95$ K) it is extremely difficult to measure them directly. 
Even in the presence of these difficulties, there have been many novel proposals to detect CNB neutrinos. Based on the proposal by Weinberg~\cite{Weinberg:1962zza}, the Princeton Tritium Observatory for Light, Early-Universe, Massive-Neutrino Yield (PTOLEMY) experiment strives to detect CNB neutrinos by capturing electron neutrinos on a 100 g tritium target via the process $^3$H + $\nu_e \to ^3$He$^{+} + e^-$~\cite{PTOLEMY:2018jst}. Apart from this, other proposals include, Stodolsky effect~\cite{Stodolsky:1974aq, Duda:2001hd}, usage of different types of interferometer~\cite{Domcke:2017aqj, Nugroho:2023cun}, bremsstrahlung processes~\cite{Asteriadis:2022zmo}, annual modulation due to gravitational focusing by the sun~\cite{Safdi:2014rza}, coherent neutrino scattering~\cite{Opher:1974drq, Lewis:1979mu, 1982ZhPmR36224S, Shergold:2021evs}, resonant scattering with cosmogenic neutrinos~\cite{Brdar:2022kpu}, atomic de-excitation~\cite{Yoshimura:2014hfa}, accelerator experiments~\cite{Bauer:2021uyj}, spectral lines from possible neutrino decays~\cite{Bernal:2021ylz} etc. For a current review on the bounds on CNB, see~\cite{Bauer:2022lri, Dey:2024agv}. Here, we are going to focus exclusively on the neutrino capture process which is the basis for the PTOLEMY experiment.
From the experimental point of view, the CNB neutrino capture process ideally can show up as a peak in the final state electron spectrum at an energy twice the mass of the neutrino, $2m_{\nu}$, above the $\beta$-decay endpoint. However, this calls for an ambitious energy resolution of $\mathcal{O}(10~\text{meV})$ if the lightest neutrino mass is in the ballpark of a few tens of meV such that it can be distinguished from the $\beta$-decay background. 
From the theoretical perspective, with the optimism that the lightest neutrino mass and the energy resolution of the detector will be in such a range that CNB can be detected, we can expect to exploit such experiments for the investigation of other theoretical concepts. For example, as an excellent spin-off of the direct measurement of CNB is the ability to distinguish between Dirac and Majorana nature of neutrinos. The neutrino capture rate is twice in the Majorana case as compared to the Dirac case~\cite{Long:2014zva}. This assertion can be modified, for example, in BSM scenarios~\cite{Zhang:2015wua, Arteaga:2017zxg}, or in the presence of a non-thermal component in the CNB~\cite{Chen:2015dka}.  It is therefore quite timely to explore various interactions dictated by BSM scenarios and study their effects in such detectors in terms of the detection rates of the CNB.   

In this work, we follow the prescription developed in \cite{Flores:2021kzl}, where the potential of neutrino scattering experiment has been examined to constrain GNIs, whereas our current focus is on testing the experimental capabilities of PTOLEMY for GNIs in a model-independent way. 
%
In~\cite{Arteaga:2017zxg} the impact of such exotic interactions on the PTOLEMY has been studied, where it was shown that the measured capture rate can not unambiguously determine the nature of neutrinos. Although it must be noted here, that the GNIs involving the vector/axial-vector operators can not be probed through only (inverse) $\beta$-decay experiments since their effects can be absorbed in the CKM matrix elements which was first observed in~\cite{Cirigliano:2012ab} and recently in \cite{Gonzalez-Alonso:2018omy, Falkowski:2019xoe}.
%
In contrast to the approach taken in ~\cite{Arteaga:2017zxg}, our analysis considers the theoretical advancements presented in \cite{Cirigliano:2012ab,Gonzalez-Alonso:2018omy, Falkowski:2019xoe} and perform a comprehensive numerical analysis.
%
Through a chi-square analysis, we systematically investigate the PTOLEMY's potential to constrain various GNI parameters in a model independent way.
We first perform  a one-parameter case, we let one GNI parameter vary while fixing all the other GNI parameters to be zero. Later, for the two-parameter analysis, we simultaneously let two GNI parameters vary while fixing the other two as zero.
Afterward, adopting benchmarks values from our chi-square analysis, we show the differential energy spectrum of the electron near the end-point
the energy of the $\beta$-decay of tritium to examine the effect of these different BSM interactions on the
spectrum, by fixing the mass of the lightest neutrino and  the energy resolution of the experiment.
Finally,  our noteworthy results are presented in the lightest neutrino mass vs GNI parameter plane.

The remainder of the article is structured as follows. In section~\ref{sec:theory_cnb} we briefly describe the basics of cosmic neutrino background for completeness. In section~\ref{sec:detection_cnb} we elaborate on the detection technique and analytical estimations at the PTOLEMY detector. Discussion of relevant GNI is given in section~\ref{sec:gni}. In section~\ref{sec:NumAnal} we explain the numerical analysis using single and double GNI parameters respectively. The electron spectrum resulting in the presence of GNI is discussed in section~\ref{sec:results}. Lastly, we summarize our findings and conclude in section~\ref{sec:concl}.

\section{Theory of Cosmic Neutrino Background}
\label{sec:theory_cnb}
For completeness, in this section, we briefly discuss some of the basic tenets of CNB. At the very early hot and dense stage of the Universe, neutrinos were prohibited from free-streaming since they maintained an equilibrium with the SM thermal bath which was constituted mostly of electrons, positrons, and photons,  via the weak interactions.  These interactions were various scattering and annihilation processes, e.g.,  $e^{-}e^{+} \rightleftharpoons \nu_{j}\bar{\nu}_{j}, ~e^{\pm}\nu_{j} \rightleftharpoons e^{\pm}\nu_{j}, ~ e^{\pm}\bar{\nu}_{j} \rightleftharpoons e^{\pm}\bar{\nu}_{j}$.  The cross sections, $\sigma$, for these processes can be exactly calculated using the standard theory of weak interaction.  A quick order-of-magnitude estimation of $\sigma$ can be easily obtained by noting the interaction strength, weak mediators, etc. as, $\sigma \sim \alpha^2 E_{\nu}^2/m_{W,Z}^4$, where $\alpha$ is the fine structure constant, $m_{W,Z}$ is the mass of the weak gauge bosons and $E_\nu$ is the neutrino energy. The relevant interaction rate can be estimated from the thermal average of the cross-section, $\langle \sigma v\rangle$ and the number density of the neutrinos, i.e., $\Gamma_{\rm int} = n \langle \sigma v\rangle$. Since at the concerned temperatures, neutrinos behave like ultra-relativistic particle species, the standard statistical mechanical calculations show the number density of neutrinos per degree of freedom to be,
\begin{align}
\label{eq:neutrino_no_density}
n = \frac{3\zeta(3)}{4\pi^2}T^3.
\end{align}
Noting that neutrinos travel at the velocity of light $v\sim c$ the typical neutrino energy is $E_\nu \sim T$, then one can estimate the interaction rate as
\begin{equation}
\label{eq:interaction_rate}
\Gamma_{\rm int}=n\langle\sigma v\rangle \sim \frac{\alpha^2}{m^4_{W,Z} }T^5 \sim G_F^2 T^5.
\end{equation}
where $G_F \approx 1.2 \times 10^{-5} \mathrm{GeV}^{-2}$ is the Fermi constant. 
As the Universe expands and cools down, the neutrinos decouple from the plasma. The expansion rate of the Universe is parametrized by the Hubble parameter $H = \dot{a}/a$, where $a$ is the scale factor. Note that $H$ at the radiation-dominated epoch ($\rho \propto T^4$) is given by,
\begin{equation}
\label{eq:hubble_parameter}
H=\frac{\dot{a}}{a}=\sqrt{\frac{8 \pi G}{3} \rho} \sim \sqrt{\frac{8 \pi G}{3} g_* \frac{\pi^2}{30} T^4} \sim \sqrt{g_*} \frac{T^2}{m_{\rm Pl}} \propto T^2 \propto a^{-2},
\end{equation}
where $G$ is the universal gravitational constant which in natural units defines the Planck mass scale, 
\begin{equation}
m_{\mathrm{Pl}}=G^{-1 / 2}=1.22 \times 10^{19} \, \mathrm{GeV},
\end{equation}
and $g_*$ is the number of relativistic degrees of freedom which at the time gets contribution only from the photons, electrons, positrons, and neutrinos:
\begin{align}
\label{eq:g*_definition}
g_*(T) = \sum_{\rm B} g_{b}+\frac{7}{8} \sum_{\rm F} g_{f}
= 2 + 4 \times \frac{7}{8}+ 6 \times \frac{7}{8} = 10.75.
\end{align}
The neutrino decouples from the thermal plasma when the interaction rate becomes less than or equal to the expansion rate, i.e., $\Gamma_{\rm int} \lesssim H$. Thus from Eqs.~\eqref{eq:interaction_rate} and~\eqref{eq:hubble_parameter}, one can get the freeze-out temperature for neutrinos $T_{\nu,f}$ to be,
\begin{gather}
G_{\mathrm{F}}^2 T_{\nu, \mathrm{f}}^5=\sqrt{g_*} \frac{T_{\mathrm{\nu,f}}^2}{m_{\mathrm{Pl}}} \Rightarrow T_{\nu, \mathrm{f}}=\left(\frac{\sqrt{g_*}}{G_{\mathrm{F}}^2 m_{\mathrm{Pl}}}\right)^{1 / 3} \sim g_*^{1 / 6} \, \mathrm{MeV} ,\notag \\
T_{\nu,\mathrm{f}}  \sim 1.48 \,  \mathrm{MeV}.
\end{gather}
Note that a more refined calculation predicts $T_{\nu_e,\mathrm{f}}  \sim 2.3 $ MeV, and $T_{\nu_\mu, \nu_\tau,\mathrm{f}}  \sim 3.5 $ MeV~\cite{Dicus:1982bz, Enqvist:1991gx}, see also~\cite{Dodelson:1992km, Hannestad:1996ui, Akita:2020szl}. From this one can get the present effective neutrino temperature, $ T_{\nu, 0}=T_{\nu, \mathrm{f}} a\left(t_\nu\right)/a\left(t_0\right) = T_{\nu, \mathrm{f}}/(1+z_\nu)$, where $z_\nu$ is the redshift when neutrinos decoupled. 
The neutrino temperature was the same as the photon temperature at the time of freeze-out, after which its temperature started to decrease as the universe kept on expanding. But after the neutrino freeze-out the thermal bath effectively was filled with electrons, positrons, and photons, of which the electrons and the positrons started to annihilate to photons as soon as the temperature dropped below the electron mass, thereby injecting entropy into the photons. 
Due to this, the rate of decrease in photon temperature slows down in comparison to the neutrino temperature as the Universe expands.
%
%
Using the fact that the entropy per co-moving volume is constant and $T_\nu \sim a^{-1}$, we can calculate the temperature of relic neutrinos as of today. Note that
\begin{equation}
s=g_* \frac{2 \pi^2}{45} T^3;\quad s_{e^{\pm}, \gamma} a^3 = \text{const.} ; \quad g_*(T) \frac{T^3}{T_\nu^3}= \text{const.}
\end{equation}
For $T \gg m_e$, electrons and positrons are still relativistic, and
$
g^{e^{\pm,\gamma}}_*(T_{\nu,\mathrm{f}}) = 2+7\times(2+2)/8=11/2.
$
After the $e^+ e^-$ annihilation, photons contribute to the entire entropy
$
g^\gamma_*(T_{\gamma,0}) \rightarrow 2 .
$
For temperatures well below electron mass, including the current epoch, we get
\begin{equation}
\frac{T_{\gamma, 0}}{T_{\nu, 0}}=\left(\frac{g^{e^{\pm,\gamma}}_*\left(T_{\nu, \mathrm{f}}\right)}{g_*^{ \gamma}\left(T_{\gamma,0}\right)}\right)^{1 / 3}=\left(\frac{11}{4}\right)^{1 / 3} \simeq 1.4 .
\end{equation}
If we use the known value of the CMB temperature today, $T_{\gamma,0}=2.725$ K~\cite{Fixsen:2009ug}, we see that neutrinos are as cold as
$
T_{\nu, 0}=1.945 \mathrm{~K}, \; \text{or}\; 1.676 \times 10^{-4} \; \mathrm{eV}.
$
Using the present temperature of neutrinos in Eq.~\eqref{eq:neutrino_no_density}, we calculate the number density of neutrinos per degree of freedom to be 
$n_{\nu, 0}=3 \zeta(3) T_{\nu,0}^3/4\pi^2  \simeq 56 \mathrm{~cm}^{-3}$.
%

%
Before proceeding further, we make a few remarks about the issue of clustering of CNB.  Since neutrinos have some tiny masses they can not escape gravitational effects and, in principle, can be trapped by the gravitational potential wells of galaxies or their clusters if the CNB neutrinos have velocities smaller than the escape velocity~\cite{Ringwald:2004np}. This may lead to a local overdensity of neutrinos and the standard density of 56 cm$^{-3}$ can be enhanced~\cite{Ringwald:2004np, Mertsch:2019qjv}. However, this is at the level that even a few years running of PTOLEMY will not be able to measure the overdensity but can put stringent bounds for certain mass ranges of the lightest neutrinos~\cite{Long:2014zva}.

\section{Detecting Cosmic Neutrino Background at PTOLEMY}
\label{sec:detection_cnb}
The current temperature of the relic neutrinos of CNB is $ T_{\nu, 0}=1.945 \mathrm{~K}, \; \text{or}\; 1.676 \times 10^{-4} \; \mathrm{eV}$ which is why they do not possess enough threshold energy to be detected in the traditional neutrino detection experiments. An interesting proposal to detect these relic neutrinos is the process of neutrino capture on the $\beta$-unstable nuclei~\cite{Weinberg:1972kfs,Cocco:2009rh,Cocco:2007za}. The greatest advantage of this process is that it requires no threshold energy of the initial state neutrinos.
Cashing in on this idea, the PTOLEMY~\cite{PTOLEMY:2019hkd} experiment is proposed to use tritium (${}^3\mathrm{H}$) as the target element since it provides the best chances of CNB detection due to its lifetime, availability, a low $Q$ value and a high neutrino capture cross-section~\cite{Cocco:2007za}. The process of neutrino capture on ${}^3\mathrm{H}$ is given by,
\begin{equation}
\nu_e + {}^3\mathrm{H} \rightarrow {}^3\mathrm{He} + e^-.
\end{equation} 
%
Recall that in the standard theory of neutrino oscillations,
a  flavour eigenstate of neutrino ($|\nu_{\alpha}\rangle$) is expressed as 
\begin{equation}
|\nu_{\alpha}\rangle=\sum_{j=1}^{3}U_{\alpha j}^{*}|\nu_i\rangle \;,
\end{equation}
where, $|\nu_i\rangle$ represent the mass eigenstates, $U$ is the mixing matrix, and in general  $\alpha$ can be $e, \mu $ or $ \tau$ corresponding to the three flavors of neutrino.
Since here we are only concerned about $\nu_e$ the relevant elements of the mixing matrix are $U_{ei}$. As the relic neutrinos propagate through the universe since the decoupling, they eventually (within $t\sim H_{0}^{-1}$) decompose into the mass eigenstates~\cite{Weiler:1999ny}. The capture rate of CNB neutrinos by ${}^3\mathrm{H}$ can then be given as,
\begin{equation}
\Gamma_{\mathrm{CNB}}=\sum_{j=1}^{3}\Gamma_j\;,
\end{equation}
$\Gamma_j$ corresponds to the capture rate of the $j$-th mass eigenstate of neutrino, which can in turn be expressed as~\cite{PTOLEMY:2019hkd},
\begin{equation}
\label{eq:CNB-Rate}
\Gamma_j = N_H\mid{U_{ej}}\mid^2\int\dfrac{d^3p_j}{(2\pi^3)}\sigma(p_j)v_jf_{j}(p_j) \;.
\end{equation}
Here, $N_H=M_H/m_{\mathrm{H}}$ is the number of tritium nuclei in a target mass of $M_H$, $p_j$ represents the neutrino momentum, $v_j$ is the neutrino velocity as measured at the detector, $\sigma(p_j)$ is the momentum dependent cross-section and $f_j(p_j)$ is the momentum distribution function of the $j$-th neutrino mass eigenstate. Because of the narrow phase space distribution of the CNB neutrinos the above integral reduces to,
\begin{equation}
\label{eq:CNB-Rate2}
\Gamma_{j} = N_H\sigma_{j}^{\mathrm{SM}}v_{j}f_{c,j}n_{\nu,0}\;,
\end{equation}  
where $f_{c,j}$ is the neutrino clustering factor due to gravitational attraction of the galactic contents~\cite{Ringwald:2004np, deSalas:2017wtt, Zhang:2017ljh,Mertsch:2019qjv}. 
Here, $\sigma_{j}^{\mathrm{SM}}$ is the average cross-section for neutrino capture according to the SM, i.e.,
\begin{equation}
\label{eq:AvgCS}
\sigma_j^{\mathrm{SM}} v_j=\frac{G_F^2}{2 \pi}\left|V_{u d}\right|^2\left|U_{e j}\right|^2 F_Z\left(E_e\right) \frac{m_{\mathrm{He}}}{m_{\mathrm{H}}} E_e p_e [g_V^2+3g_A^2]\;,
\end{equation}
where, $E_e$ is the electron energy, $p_e$ is the electron momentum, $m_{\mathrm{He}}\approx 2808.391\mathrm{~MeV}$ and $m_{\mathrm{H}}\approx 2808.921\mathrm{~MeV}$. The quantities $g_V$ and $g_A$ are the concerned vector and axial-vector couplings. The Fermi function, $F_Z\left(E_e\right)$ takes into account the Coulomb interaction between a proton and an outgoing electron. This is can be approximately expressed as~\cite{Primakoff:1959chj},
\begin{equation}
F_Z\left(E_e\right)=\dfrac{2\pi Z\alpha E_e/p_e}{1-\exp(-2\pi Z\alpha E_e/p_e)}\;.
\end{equation}
In our concerned case, $Z=2$ is the atomic number of ${}^3\mathrm{He}$ and $\alpha=1/137.036$ is the fine structure constant.
In the expressions of the capture rate  and cross-section  (see Eqs.~\eqref{eq:CNB-Rate} and \eqref{eq:AvgCS}), the mixing parameter $\left|U_{ej}\right|^2$ appears because even though the neutrinos propagate in their mass eigenstates they are expected to be captured in their flavor eigenstate, i.e. $\nu_{e}$. 
One of the most crucial background of CNB neutrino capture is the electrons of the highest energy produced from the $\beta$-decay of tritium itself. To take these $\beta$-decays into account, we express the $\beta$ decay spectrum as~\cite{Masood:2007rc,  PTOLEMY:2019hkd},
\begin{equation}
\dfrac{d\Gamma_{\beta}}{dE_e}=\dfrac{N_H}{\pi^2}\sum_{j=1}^{3}\sigma_j^{\mathrm{SM}}H(E_e,m_j) \;,
\end{equation}
where,
\begin{equation}
H(E_e,m_j)=\dfrac{1-m_e^2/(E_em_{\mathrm{H}})}{(1-2E_e/m_{\mathrm{H}}+m_e^2/m_{\mathrm{H}}^2)^2}\sqrt{y\left(y+\dfrac{2m_jm_{\mathrm{He}}}{m_{\mathrm{H}}}\right)}\left[y+\dfrac{m_j}{m_{\mathrm{H}}}\left(m_{\mathrm{He}}+m_j\right)\right]\;,
\end{equation}
for $y=E_{\mathrm{end},0}-E_e-m_j$, and $E_{\mathrm{end},0}$ is the $\beta$-decay endpoint energy for neutrinos with zero mass. 
%
%
It is noteworthy that recent studies have determined the realistic experimental resolution of PTOLEMY, revealing that, under the existing design, PTOLEMY will be facing challenges to reach the desire goal due to the quantum effect of the zero-point motion of adsorbed tritium~\cite{Cheipesh:2021fmg}. Nevertheless, endeavors are underway to enhance the resolution by modifying the experiment's design~\cite{PTOLEMY:2022ldz}. In our analysis, we adopt a benchmark value of $\Delta=20\mathrm{~meV}$, which could potentially be achievable in the future with design modifications.

%
It must be noted that the experimental energy resolution $\Delta$ plays a pivotal role in the detection of the CNB. The effect of $\Delta$ can be incorporated into the calculation by convoluting both the CNB and $\beta$-decay part of the electron spectrum with a Gaussian of full width at half maximum given by $\Delta$, which in turn smears the electron spectrum.
The smeared $\beta$-decay spectrum is expressed as,
\begin{equation}
\label{eq:Beta-spectrum}
\dfrac{d\bar{\Gamma}_{\beta}}{dE_e}(E_e)=\dfrac{1}{\sqrt{2\pi}(\Delta/\sqrt{8\mathrm{ln}2})}\int^{+\infty}_{-\infty}dE'\dfrac{d\Gamma_{\beta}}{dE_e}(E')\mathrm{exp}\left[-\dfrac{(E_e-E')^2}{2(\Delta/\sqrt{8\mathrm{ln}2})^2}\right]\;.
\end{equation}
Similarly, the smeared CNB neutrino capture rate can be expressed as,
\begin{equation}
\label{eq:CNB-spectrum}
\dfrac{d\bar{\Gamma}_{\mathrm{CNB}}}{dE_e}(E_e)=\dfrac{1}{\sqrt{2\pi}(\Delta/\sqrt{8\mathrm{ln}2})}\sum_{j=1}^{3}\Gamma_j\mathrm{exp}\left[-\dfrac{[E_e-(E_{\mathrm{end}}+m_j+m_{\mathrm{lightest}})]^2}{2(\Delta/\sqrt{8\mathrm{ln}2})^2}\right]\;,
\end{equation}
where $E_{\mathrm{end}}=E_{\mathrm{end},0}-m_{\mathrm{lightest}}$ is the endpoint energy of the $\beta$-decay, $m_{\mathrm{lightest}}$ being the mass of the lightest neutrino and $\Gamma_j$ is expressed by Eq.~\eqref{eq:CNB-Rate2}.
These expressions for the smeared electron spectrum will be used to perform the statistical analysis relevant to our study in the subsequent sections of the article.

\section{Generalized Interactions of Relic Neutrinos}
\label{sec:gni}
In this section we briefly discuss the generalized neutrino interactions (GNI), the effect of which we are going to study at the PTOLEMY.
%

%
%

Besides the usual SM interactions, Wolfenstein in \cite{Wolfenstein:1977ue}  first proposed that  interactions of neutrinos with matter can lead to neutrino non-standard interactions (NSI) -- a type of \textit{new physics} interactions beyond just the mass generation -- which may appear in unknown couplings.
The effective Lagrangian of such new interactions is usually expressed in (chiral) vector form and is given by,
\begin{equation}\label{eq:NSI}
  \mathcal{L}^{\rm NC/CC}_\text{NSI} \supset  \dfrac{G_F}{\sqrt{2}} \epsilon^{ff^{\prime}  }_{\alpha\beta}  \left[ \overline{\nu}_\alpha \gamma^{\rho} (1 - \gamma^{5}) \nu_\beta \right] 
\left[ \bar{f} \gamma_{\rho}  (1 \mp \gamma^{5}) f^{\prime}\right] 
  + \text{h.c.} \;,
\end{equation}
where $ \epsilon^{ff^{\prime}}_{\alpha\beta}$ represent the NSI parameters,  $  f,  f^{\prime} = e, u, d$,  and  $ \alpha, \beta = e, \mu, \tau $.  For $ f = f^{\prime} $ the NSIs are neutral-current (NC) like, else it is charged-current (CC) like.
%

%

Here, we aim to investigate the most general Lorentz-invariant interactions beyond the usual chiral and vector-like form of the NSIs for SM neutrinos. These exotic neutrino interactions  are  called \textit{generalized neutrino interactions} (GNI) \cite{Lindner:2016wff}. 
Since we are interested to study the effect of GNIs in relic neutrino capture on $\beta$-decaying tritium, expressing the effective Lagrangian in mass eigenstates is more appropriate in order to obtain the relevant interactions,
\begin{equation}
\mathscr {L}_{\mathrm{eff}}=-\dfrac{G_F}{\sqrt{2}}V_{ud}U_{ej}\left\lbrace[\bar{e}\gamma^{\mu}(1-\gamma^5)\nu_j][\bar{u}\gamma_{\mu}(1-\gamma^5)d]+\sum_{l,q}\epsilon_{lq}[\bar{e}\mathcal{O}_l\nu_j][\bar{u}\mathcal{O}_qd]\right\rbrace+\mathrm{h.c.} \;,
\label{eq:Lagrangianeff}
\end{equation}
where the dimensionless couplings $\epsilon_{lq}$ represent the GNI parameters and $j = 1, 2, 3$ are the three  mass eigenstates of neutrino. The operators $\mathcal{O}_l$ and $\mathcal{O}_q$ are the relevant lepton and quark current, respectively, and are  given in Tab.~\ref{table:2}. The quantities $V_{ud}$ and $U_{ej}$ are the relevant elements of the Cabibbo-Kobayashi-Maskawa (CKM) and Pontecorvo-Maki-Nakagawa-Sakata (PMNS) matrices, respectively. 
%
%
In this study, we have conducted an analysis considering both left- and right-handed neutrinos in a model-independent manner. This approach allows us to explore scenarios where neutrino mass can be either of Dirac or Majorana nature.
%
%
%
%
\begin{table}[h!]
\begin{center}
\begin{tabular}{|| c c c ||}
\hline
$\epsilon_{lq}$ & $\mathcal{O}_{l}$ & $\mathcal{O}_q$ \\ [0.5ex]
\hline
\hline 
$\epsilon_{LL}$ & $\gamma^{\mu}(1-\gamma^5)$ & $\gamma_{\mu}(1-\gamma^5)$\\ \hline
$\epsilon_{LR}$ & $\gamma^{\mu}(1-\gamma^5)$ & $\gamma_{\mu}(1+\gamma^5)$\\ \hline
$\epsilon_{RL}$ & $\gamma^{\mu}(1+\gamma^5)$ & $\gamma_{\mu}(1-\gamma^5)$\\ \hline
$\epsilon_{RR}$ & $\gamma^{\mu}(1+\gamma^5)$ & $\gamma_{\mu}(1+\gamma^5)$\\ \hline
$\epsilon_{LS}$ & $1-\gamma^5$ & $1$\\ \hline
$\epsilon_{RS}$ & $1+\gamma^5$ & $1$\\ \hline
$\epsilon_{LT}$ & $\sigma^{\mu\nu}(1-\gamma^5)$ & $\sigma_{\mu\nu}(1-\gamma^5)$\\ \hline
$\epsilon_{RT}$ & $\sigma^{\mu\nu}(1+\gamma^5)$ & $\sigma_{\mu\nu}(1+\gamma^5)$\\[1ex]
\hline
\end{tabular}
\caption{GNI parameters and the Lorentz structures corresponding to them relevant in this study.}
\label{table:2}
\end{center} 
\end{table}
Using the effective Lagrangian given in Eq.~\eqref{eq:Lagrangianeff} we analyze the neutrino absorption on tritium following the steps mentioned in the previous section. We can express the relevant hadronic matrix elements corresponding to the quark current mentioned in Eq.~\eqref{eq:Lagrangianeff} and Tab.~\ref{table:2} as~\cite{Weinberg:1958ut, Cirigliano:2013xha, Ludl:2016ane},
\begin{subequations}
\begin{align}
\left\langle p\left(p_p\right)|\bar{u} d| n\left(p_n\right)\right\rangle &=g_S\left(q^2\right) \overline{u_p}\left(p_p\right) u_n\left(p_n\right) \;, \\
\left\langle p\left(p_p\right)\left|\bar{u} \sigma^{\mu \nu}\left(1 \pm \gamma^5\right) d\right| n\left(p_n\right)\right\rangle &=g_T\left(q^2\right) \overline{u_p}\left(p_p\right) \sigma^{\mu \nu}\left(1 \pm \gamma^5\right) u_n\left(p_n\right) \;,\\
\left\langle p\left(p_p\right)\left|\bar{u} \gamma^\mu\left(1 \pm \gamma^5\right) d\right| n\left(p_n\right)\right\rangle &=\overline{u_p}\left(p_p\right) \gamma^\mu\left[g_V\left(q^2\right) \pm g_A\left(q^2\right) \gamma^5\right] u_n\left(p_n\right) \;.
\end{align}
\label{eq:hadronicformfactors}
\end{subequations} 
Though these form-factors depend on the transferred momentum $q^2=(p_n-p_p)^2$, for the capture rate of the non-relativistic CNB neutrinos one can safely take the limit $q^2\sim 0$. Here, $g_V$, $g_A$, $g_S$, and $g_T$ correspond to the form-factors of vector, axial-vector, pseudo-scalar and tensor Lorentz structures. The values of the form-factors used in our analysis is given in Tab.~\ref{table:3}.
\begin{table}[h!]
\begin{center}
\begin{tabular}{|| c c c ||}
\hline
Form Factor & Value & Reference \\ [0.5ex]
\hline
\hline 
$g_V(0)$ & $1$ & \cite{Gershtein:1955fb, Feynman:1958ty} \\ \hline
$\tilde{g}_A(0)/g_V(0)$ & $1.278\pm 0.0021$ & \cite{Berkowitz:2017gql} \\ \hline
$g_S(0)$ & $1.02\pm 0.11$ & \cite{Gonzalez-Alonso:2013ura} \\ \hline
$g_T(0)$ & $1.020\pm 0.076$ & \cite{Bhattacharya:2015esa}\\[1ex]
\hline
\end{tabular}
\caption{\footnotesize Values of the hadronic form factors used in this work.}
\label{table:3}
\end{center} 
\end{table} 
%
%
%
The relic neutrino capture cross-section in the presence of GNI for a neutrino mass-eigenstate $j$, with helicity $h_j=\pm 1$ and velocity $v_j$ can be expressed as \cite{Long:2014zva},
\begin{equation}
\sigma_j^{\mathrm{BSM}}\left(h_j\right) v_j=\frac{G_F^2}{2 \pi}\left|V_{u d}\right|^2\left|U_{e j}\right|^2 F_Z\left(E_e\right) \frac{m_{\mathrm{He}}}{m_{{\mathrm{H}}}} E_e p_e \mathcal{M}_j\left( \epsilon_{l q}\right) \;,
\label{eq:crosssection}
\end{equation}
where $\mathcal{M}_j(\epsilon_{lq})$ is the amplitude-squared and it carries the information of the GNI parameters $\epsilon_{lq}$. 
We like to point out that all GNI terms tabulated in table~\ref{table:2} cannot be tested using only (inverse) $\beta$-decay  experiments. This is due to the fact that the vector and axial-vector GNIs come along with SM couplings and hence can be absorbed in the CKM matrix elements.
This can easily be demonstrated by expanding  Eq.~\eqref{eq:Lagrangianeff}, keeping only vector and axial vector couplings~\cite{Herczeg:2001vk,Cirigliano:2012ab},
\begin{align}
\label{eq:LagrangianeffVA}
\mathscr{L}_{\mathrm{eff}}^{V,A} = 
 -\dfrac{G_F}{\sqrt{2}}V_{ud}U_{ej}
 \bigg[
 (1 &+ \epsilon_{LL})
 [\bar{e}\gamma^{\mu}(1-\gamma^5)\nu_j]
 [\bar{u}\gamma_{\mu}(1-\gamma^5)d] 
 \nonumber \\
& +\epsilon_{LR}[\bar{e}\gamma^{\mu}(1-\gamma^5)\nu_j]
[\bar{u}\gamma_{\mu}(1+\gamma^5)d] 
\nonumber \\
& +\epsilon_{RL}[\bar{e}\gamma^{\mu}(1+\gamma^5)\nu_j][\bar{u}\gamma_{\mu}(1-\gamma^5)d]
\nonumber \\
 &+\epsilon_{RR}[\bar{e}\gamma^{\mu}(1+\gamma^5)\nu_j][\bar{u}\gamma_{\mu}(1+\gamma^5)d]\bigg]+\mathrm{h.c.}
\end{align}
Therefore, in the presence of such GNIs $V_{ud}$ can be redefined as,
\begin{align}
\left|\tilde{V}_{ud}\right|^2\approx\left|V_{ud}\right|^2(1+\epsilon_{LL}+\epsilon_{LR}+\epsilon_{RL}+\epsilon_{RR})^2 \;.
\label{eq:vudtilde}
\end{align} 
Similarly, since form factor of the axial vector Lorentz structure, $g_A$ is the phenomenological value obtained from the $\beta$ asymmetry parameter, it will also have the GNI parameters measured along with it, i.e.
\begin{align}
\tilde{g}_A\approx g_A\dfrac{1+\epsilon_{LL}-\epsilon_{LR}+\epsilon_{RR}-\epsilon_{RL}}{1+\epsilon_{LL}+\epsilon_{LR}+\epsilon_{RR}+\epsilon_{RL}} \;.
\end{align} 
Taking these experimental inputs into consideration, we find that Eq.~\eqref{eq:crosssection} takes the form,
\begin{align}
\sigma_j^{\mathrm{BSM}}\left(h_j\right) v_j=\frac{G_F^2}{2 \pi}\left|\tilde{V}_{u d}\right|^2\left|U_{e j}\right|^2 F_Z\left(E_e\right) \frac{m_{\mathrm{He}}}{m_{{\mathrm{H}}}} E_e p_e \tilde{\mathcal{M}}_j\left(\epsilon_{l q}\right) \;,
\label{eq:crosssectionmod}
\end{align}
and $\tilde{\mathcal{M}}_j\left(\epsilon_{l q}\right)$ is given by,
\begin{footnotesize}
\begin{align}
\tilde{\mathcal{M}}_j\left(\epsilon_{l q}\right)&=\dfrac{g_V^2}{\mathcal{D}_1^2} \left((1+\epsilon_{LL}+\epsilon_{LR})^2+(\epsilon_{RR}+\epsilon_{RL})^2\right)+g_S^2\left(\epsilon_{LS}^2+\epsilon_{RS}^2\right)+48 g_T^2\left(\epsilon_{LT}^2+\epsilon_{RT}^2\right) \nonumber \\
& ~~ +\dfrac{3\tilde{g}_A^2}{\mathcal{D}_2^2}\left((1+\epsilon_{LL}-\epsilon_{LR})^2+(\epsilon_{RR}-\epsilon_{RL})^2\right) \nonumber \\
& ~~ +\dfrac{2 g_S g_V}{\mathcal{D}_1^2}\left[\dfrac{m_e}{E_e}\left(\epsilon_{LS}(1+\epsilon_{LL}+\epsilon_{LR})+\epsilon_{RS}(\epsilon_{RR}+\epsilon_{RL})\right)+\left(\epsilon_{RS}(1+\epsilon_{LL}+\epsilon_{LR})+\epsilon_{LS}(\epsilon_{RR}+\epsilon_{RL})\right)\right] \nonumber \\
& ~~-\dfrac{24 \tilde{g}_A g_T}{\mathcal{D}_1\mathcal{D}_2}\left[\dfrac{m_e}{E_e}\left(\epsilon_{LT}(1+\epsilon_{LL}-\epsilon_{LR})+\epsilon_{RT}(\epsilon_{RR}-\epsilon_{RL})\right)+\left(\epsilon_{RT}(1+\epsilon_{LL}-\epsilon_{LR})+\epsilon_{LT}(\epsilon_{RR}-\epsilon_{RL})\right)\right] \nonumber \\
& ~~+\dfrac{2 m_e}{E_e \mathcal{D}_1^2}\left[g_V^2(1+\epsilon_{LL}+\epsilon_{LR})(\epsilon_{RR}+\epsilon_{RL})+g_S^2 \epsilon_{RS}\epsilon_{LS}+48 g_T^2\epsilon_{LT}\epsilon_{RT}\right] \nonumber \\
& ~~+\dfrac{2 m_e}{E_e \mathcal{D}_2^2}3 \tilde{g}_A^2(1+\epsilon_{LL}-\epsilon_{LR})(\epsilon_{RR}-\epsilon_{RL}) \;,
\label{eq:BSMamplitude}
\end{align}
\end{footnotesize}
where $\mathcal{D}_1=(1+\epsilon_{LL}+\epsilon_{LR}+\epsilon_{RR}+\epsilon_{RL})$ and $\mathcal{D}_2=(1+\epsilon_{LL}-\epsilon_{LR}+\epsilon_{RR}-\epsilon_{RL})$. Due to the non-relativistic nature of the relic neutrinos, we have taken $E_j\sim m_j$.
To obtain Eq.~\eqref{eq:BSMamplitude}, we have used hadronic matrix elements of Eqs.~\eqref{eq:hadronicformfactors} in Eq.~\eqref{eq:Lagrangianeff}.
Note that here we take into account the effect of the GNI parameters on $g_A$ and $V_{ud}$ in constructing the amplitude-squared which was ignored in~\cite{Arteaga:2017zxg} resulting in an imprecise estimation of those GNI parameters in the numerical analysis.
Due to the presence of $\mathcal{D}_1$ and $\mathcal{D}_2$ in the above expression, probing $\epsilon_{LL}$, $\epsilon_{LR}$, $\epsilon_{RL}$, and $\epsilon_{RR}$ becomes impossible. Even in the case of interference between any of these vector interactions with scalar or tensor interactions, the vector couplings just scale the tensor or scalar couplings. 
In order to illustrate this further, we take the specific examples of the presence of only $\epsilon_{LS}$ and later both non-zero $\epsilon_{LS}$ and $\epsilon_{LL}$. Using Eq.~\eqref{eq:BSMamplitude}, we find that
\begin{align}
\tilde{M}_j(\epsilon_{LS})=g_V^2+3\tilde{g}_A^2+g_S^2\epsilon_{LS}^2+\dfrac{2 m_e}{E_e}g_Sg_V\epsilon_{LS} \;,
\end{align}
and,
\begin{align}
\tilde{M}_j(\epsilon_{LS},\epsilon_{LL})=g_V^2+3\tilde{g}_A^2+g_S^2\dfrac{\epsilon_{LS}^2}{(1+\epsilon_{LL})^2}+\dfrac{2 m_e}{E_e}g_Sg_V\dfrac{\epsilon_{LS}}{(1+\epsilon_{LL})} \;.
\end{align}
It can be seen from the two above expressions that the expressions are identical to one another and the presence of $\epsilon_{LL}$ just scales $\epsilon_{LS}$, i.e. $\epsilon_{LS}\rightarrow\epsilon_{LS}/(1+\epsilon_{LL})$.
Therefore, in the subsequent sections, we will consider the effect of only the scalar and tensor BSM interactions on capture cross-section in order to determine their effects on the electron spectrum due to CNB capture in a PTOLEMY or PTOLEMY-like experiment through statistical analysis.

\section{Numerical Analysis}
\label{sec:NumAnal}
In this section, we outline our methodology to characterize the sensitivity of the PTOLEMY detector in order to determine statistical bounds on the GNI parameters. We perform a $\chi^2$-analysis by adopting the technique developed in~\cite{KATRIN:2005fny}. 
%
We start our simulation by evaluating the number of $\beta$-decays ($ N_{\beta}^i $) and the relic neutrino capture events ($ N_{\mathrm{CNB}}^i $) as~\cite{PTOLEMY:2019hkd},
\begin{align}
\label{eq:Nbetai}
N_{\beta}^i &= T\int^{E_i+\delta /2}_{E_i-\delta /2}\dfrac{d\bar{\Gamma}_{\beta}}{dE_e}dE_e \;,\\
\label{eq:Ncnbi}
N_{\mathrm{CNB}}^i &= T\int^{E_i+\delta /2}_{E_i-\delta /2}\dfrac{d\bar{\Gamma}_{\mathrm{CNB}}}{dE_e}dE_e \;,
\end{align}
where `$i$' represents $i$-th bin for a given energy  $E_i$ with a bin-width $\delta$ and $T$ is the exposure time. We fix $\delta = 10$ meV. The differential capture rates are given by Eqs.~\eqref{eq:Beta-spectrum} and~\eqref{eq:CNB-spectrum}, respectively.
Here, both the event numbers  $ N_{\beta}^i $ and $N_{\mathrm{CNB}}^i $ are functions of the endpoint energy of $\beta$-decay spectrum, $E_{\mathrm{end}}$, the parameters of the leptonic mixing matrix, $U$, and the active neutrino masses, $m_i$.
In the next step, we estimate the number of expected events for a given bin to be the sum of the events mentioned in Eqs.~\eqref{eq:Nbetai} and~\eqref{eq:Ncnbi},
\begin{equation}
N^i(E_{\mathrm{end}},m_j,U_{ej}) = N_{\beta}^i(E_{\mathrm{end}},m_j,U_{ej}) + N_{\mathrm{CNB}}^i(E_{\mathrm{end}},m_j,U_{ej})  \;.
\end{equation}
Eventually, the total number of events in a given energy bin is calculated by taking into account the constant background events, $N_{\text{Bkg}}$, which can be read as
\begin{equation}
N_{\text{tot}}^i(E_{\mathrm{end}},m_j,U_{ej}) = N^i(E_{\mathrm{end}},m_j,U_{ej}) + N_{\text{Bkg}} \;.
\end{equation}
In a PTOLEMY-like experiment for the purpose of our study, the number of background events in the region around the endpoint energy is $\mathcal{O}(1)$, as we are assuming a background decay rate of $10^{-5}$ Hz in the $15$ eV energy range around the endpoint energy~\cite{PTOLEMY:2019hkd}. Next, we use the Asimov data set~\cite{Cowan:2010js} i.e.,  the dataset where there are no statistical fluctuations around the calculated number of events to estimate the experimental measurement in each energy bin,
\begin{equation}
N_{\mathrm{exp}}^i(E_{\mathrm{end}},m_j,U_{ej}) = N_{\text{tot}}^i(E_{\mathrm{end}},m_j,U_{ej}) \pm \sqrt{N_{\text{tot}}^i(E_{\mathrm{end}},m_j,U_{ej})} \;,
\end{equation}
where we have assumed a statistical error of $\sqrt{N_{\text{tot}}^i}$. 

%

%
Following a similar process to calculate the  number of events for the standard scenario, the expected number of events in the presence of GNIs can be expressed as,
\begin{equation}
N_{\text{GNI-th}}^i(E_{\mathrm{end}},m_j,U_{ej},\epsilon_{lq})= N_{\beta}^i(E_{\mathrm{end}},m_j,U_{ej},\epsilon_{lq}) +N_{\text{GNI-CNB}}^i(E_{\mathrm{end}},m_j,U_{ej},\epsilon_{lq})+N_{\text{Bkg}}\; ,
\end{equation}
where $N_{\text{GNI-CNB}}^i(E_{\text{end}},m_j,U_{ej},\epsilon_{lq})$ is the number of events in each energy bin of width $\Delta$ around the energy $E_i$ during an exposure time $T$ for relic neutrino capture in presence of GNI.
With this prescription, we define the expression of $\chi^2$ as,
\begin{equation}\label{eq:chi-sq}
\chi^2 = \sum_{i}\left(\dfrac{N^i_{\mathrm{exp}}(E_{\mathrm{end}},m_j,U_{ej})-N_{\text{GNI-th}}^i(E_{\mathrm{end}},m_j,U_{ej},\epsilon_{lq})}{\sqrt{N_{\text{tot}}^i}}\right)^2 \;.
\end{equation}
Recall that $N_{\text{GNI-th}}^{i}$ represents the theoretical  number of events in the presence of GNIs, whereas $N^i_{\text{exp}}$ means the experimental data  in the absence of new physics, i.e., just the SM expectation.
%

%
For the rest of our analysis, we parameterize the relevant leptonic mixing parameters as,
\begin{equation}
\left|U_{e1}\right|^2 = c_{12}^2c_{13}^2 \,,~~
\left|U_{e2}\right|^2 = s_{12}^2c_{13}^2 \,,~~
\left|U_{e3}\right|^2 = s_{13}^2,\;
\end{equation}
where $c_{ij}=\cos\theta_{ij}$ and $s_{ij}=\sin\theta_{ij}$, $\theta_{ij}$ being the mixing angle. In this numerical  analysis, we have used the values of mixing angles as given by Tab.~\ref{tab:OscParam}.

\begin{table}[H]
\centering
\begin{tabular}{|l|l|l|}
\hline
Parameter & Normal Ordering & Inverted Ordering \\  \hline \hline 
      $\theta_{12}({}^{\circ})$   & $33.44^{+0.77}_{-0.74}$ & $33.45^{+0.77}_{-0.74}$\\ \hline
      $\theta_{13}({}^{\circ})$   & $8.57^{+0.13}_{-0.12}$ & $8.60^{+0.12}_{-0.12}$ \\ \hline
      $\theta_{23}({}^{\circ})$   & $49.2^{+1.0}_{-1.3}$ & $49.5^{+1.0}_{-1.2}$ \\ \hline
      $\theta_{12}({}^{\circ})$   & $33.44^{+0.77}_{-0.74}$ & $33.45^{+0.77}_{-0.74}$\\ \hline
      $\Delta m_{21}^2(\mathrm{eV}^2)$   & $7.42^{+0.21}_{-0.20} \times 10^{-5}$ & $7.42^{+0.21}_{-0.20} \times 10^{-5}$\\ \hline
      $\Delta m_{3l}^2(\mathrm{eV}^2)$   & $2.515^{+0.028}_{-0.028} \times 10^{-3}$ & $-2.498^{+0.029}_{-0.029} \times 10^{-3}$\\[1ex] \hline
\end{tabular}
\caption{\footnotesize Global-fit of neutrino oscillation data~\cite{Esteban:2020cvm, nufit}. Here, the mentioned values are given as the best-fit value at $\pm 1\sigma$. Also, it is to be noted that for normal ordering, $\Delta m_{3l}^2=\Delta m_{31}^2$ and for inverted ordering $\Delta m_{3l}^2=\Delta m_{32}^2$.}
\label{tab:OscParam}
\end{table}
%

As already mentioned, in our analysis we consider bin-width to be $10\mathrm{~meV}$, i.e., $\delta=10\mathrm{~meV}$, unless otherwise stated.
%
It is noteworthy that in conducting this analysis, we have adopted an optimistic scenario by assuming an experimental resolution of $\Delta=20\mathrm{~meV}$. However, it is acknowledged that achieving this resolution with the current setup is challenging. The rationale behind this choice is to assess the limitations of PTOLEMY under the most favorable conditions. It is important to emphasize that these limitations represent upper bounds; in other words, PTOLEMY's ability to observe the effects of new interactions will be less pronounced if the experimental resolution exceeds $20\mathrm{~meV}$.
%

%
Having discussed our fitting procedure, we now perform our numerical analysis.

\subsection{One-parameter Analysis}
\begin{figure}[t]
\centering
\includegraphics[scale=0.35]{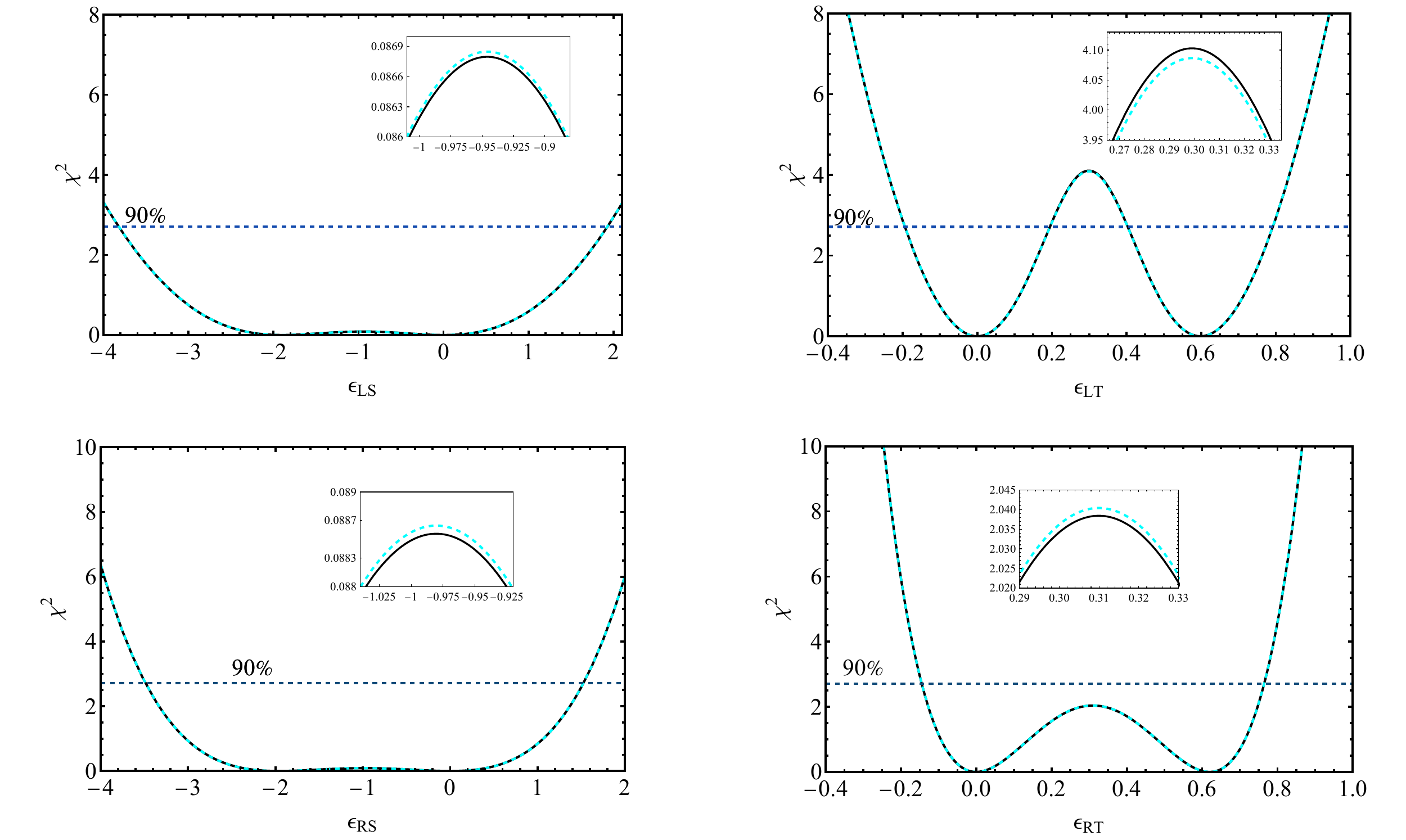}
\caption{Dependence of $\chi^2$ on $\epsilon_{lq}$ for $l = L, R$ and $q = S, T$ for different hierarchies. The solid black (dashed cyan)  line corresponds to the normal (inverted) neutrino mass ordering, whereas the horizontal dotted blue line represents the 90\% confidence level. The insets show the difference between the $\chi^2$ values for the different mass orderings.}
\label{fig:OneParam}
\end{figure}

We first perform the $\chi^2$ analysis considering one non-zero GNI parameter at a time  following  Eq.~\eqref{eq:chi-sq}. The results are presented for the normal and inverted neutrino mass ordering  in Fig.~\ref{fig:OneParam} using the solid black and dotted cyan lines, respectively.

%
By observing all the panels, it can be noted that  a value of $\epsilon_{lq} = 0$ fits the data (i.e., for no new physics contribution) in the same way, a value of $\epsilon_{lq} \neq 0 $  resembles the measured signal due to a partial destructive interference with the SM value.
For an illustrative purpose to understand this analytically, we consider Eq.~\eqref{eq:BSMamplitude} 
taking all GNI parameters to zero, except $\epsilon_{RT}$. With this assumption Eq.~\eqref{eq:BSMamplitude} simplifies to,
\begin{equation}
\tilde{\mathcal{M}}_j\left(\epsilon_{RT}\right)=g_V^2+3 \tilde{g}_A^2+48 g_T^2\epsilon_{RT}^2-24\tilde{g}_A g_T\epsilon_{RT} \;.
\label{eq:BSMamplitude2}
\end{equation}
One can notice that the right-hand side of Eq.~\eqref{eq:BSMamplitude2} leads to the same expressions $[g_V^2+3\tilde{g}_A^2]$, for $\epsilon_{RT} \sim 0 $ and/or $\tilde{g}_A/(2 g_T)\sim 0.62$. This analytical understanding resembles the numerical analysis as shown in the first panel of Fig.~\ref{fig:OneParam}, where we observe minimum values of  $\chi^2$ for $\epsilon_{RT} \sim (0, 0.62) $. 
This leads to the presence of intrinsic degeneracy for $\epsilon_{RT} $, meaning that two different sets of $\epsilon_{RT} $ lead to the same $\chi^2$ values. The PTOLEMY experiment with its future data can not remove such intrinsic degeneracy.
Likewise, the presence of degeneracy  for other GNI parameters can also be understood from the remaining panels. The values where the degeneracies occur in $\chi^2$ values for $\epsilon_{LT}$ is ($0, 0.6$).
We notice further that degeneracy is discrete for the second panel at 90\% confidence level (CL) i.e., the complete allowed parameter regions are not below the horizontal blue dashed line for $\epsilon_{LT}$. On the other hand, the allowed regions are below the horizontal blue dashed line for the other three panels, and that leads to a continuous degeneracy to $\epsilon_{RT}$, $\epsilon_{RS}$ and $\epsilon_{LS}$, respectively.
\begin{table}[h!]
\centering
\begin{tabular}{|l|l|l|}
\hline
GNI Parameter &
  \begin{tabular}[c]{@{}l@{}} $\chi^2$ values at 90\% CL \end{tabular} &\begin{tabular}[c]{@{}l@{}} $\chi^2$ values at 90\% in Ref. \cite{Escrihuela:2023sfb} \end{tabular}  \\ \hline \hline
$\epsilon_{LS}$ & $[-3.8,1.9]$ & $[0,2]$ \\ \hline
$\epsilon_{LT}$ & $[-0.19,0.19] , [0.4,0.8]$ & $[0,0.3]$\\ \hline
$\epsilon_{RS}$ & $[-3.5,1.5]$ & $-$\\ \hline
$\epsilon_{RT}$ & $[-0.15,0.75]$ & $-$\\ \hline
\end{tabular}
\caption{\footnotesize Allowed values of different  GNI parameters, $\epsilon_{QX}~(Q=L,R,~X=S,T)$, corresponding to the $90\%$ CL obtained from Fig.~\ref{fig:OneParam}. }
\label{table:table4}
\end{table}

We also observe that the PTOLEMY can put the most stringent constraints on GNI parameter $\epsilon_{LT}$ as can be seen from the top-right panel at 90\% CL. The bounds at 90\% CL are summarized in Tab.~\ref{table:table4}.
To provide a qualitative comparison of PTOLEMY's capabilities, we show available bounds for electron neutrino coming from FASER$\nu$ experiment in the third column of Tab. \ref{table:table4} of~\cite{Escrihuela:2023sfb}. In Ref.~\cite{Escrihuela:2023sfb}, only positive values of the GNI parameters relevant for left-handed neutrinos are considered. We find that in case of $\epsilon_{LS}$, our bound in the positive domain is slightly more stringent than the one from the FASER$\nu$ perspective. On the other hand, in case of $\epsilon_{LT}$, we show a degeneracy at $\epsilon_{LT}=0, 0.6$.
It is also to be noted here that in this analysis we have taken the mass of the lightest neutrino to be 50 meV as the $\chi^2$ values for different cases of GNI parameters do not depend on the mass of the lightest neutrino. This is due to the fact that, $m_{\mathrm{lightest}}$ just shifts the peak of the electron spectra due to the CNB absorption, and therefore changes the number of electrons in different energy bins. However, since we consider multiple bins, if the number of electrons in one bin is reduced due to the shifting of the peak, it increases in another bin and the overall effect is nullified.

\subsection{Two-parameter Analysis}
\begin{figure}[t!]
\includegraphics[scale=0.265]{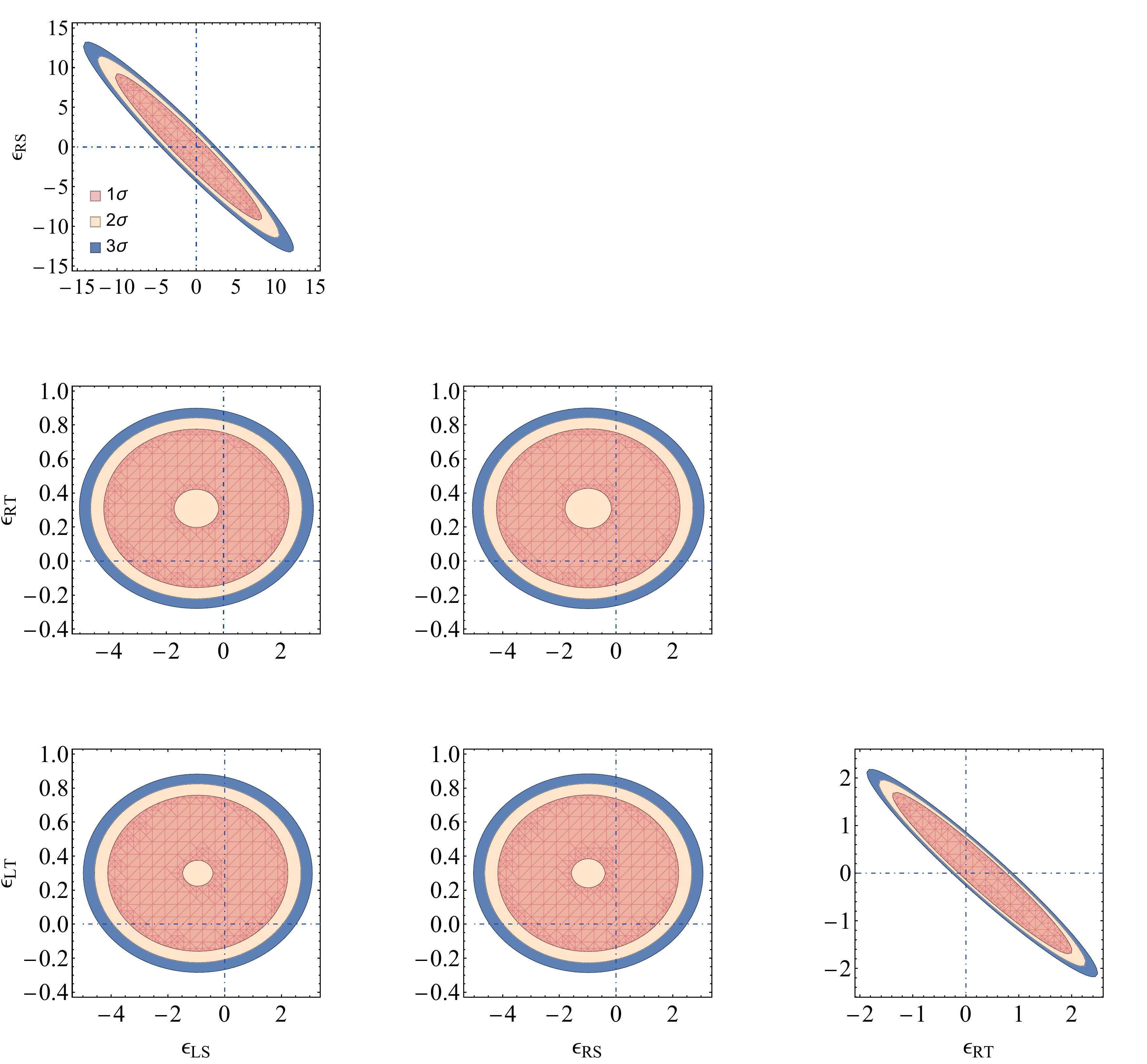}
\caption{\footnotesize Allowed region plots for two non-zero GNI parameters. In the first row ($\epsilon_{LS}, \epsilon_{RS}$), in the second row ($\epsilon_{LS},\epsilon_{RT}$) and ($\epsilon_{RS}, \epsilon_{RT}$) and in the third row ($\epsilon_{LS}, \epsilon_{LT}$), ($\epsilon_{RS}, \epsilon_{LT}$) and ($\epsilon_{RT}, \epsilon_{LT}$) have been shown. The blue dashed lines corresponds to the zero values of the different GNI parameters.}
\label{fig:TwoParam}
\end{figure}

For completeness, we perform now $\chi^2$ analysis considering two non-zero GNI parameters at a time, unlike previous section, where a single non-zero parameter has been adopted at a time. We end up with six such possibilities, i.e. $\chi^2(\epsilon_{LS},\epsilon_{LT})$, $\chi^2(\epsilon_{LS},\epsilon_{RT})$, $\chi^2(\epsilon_{LS},\epsilon_{RS})$, $\chi^2(\epsilon_{RS},\epsilon_{LT})$, $\chi^2(\epsilon_{RS},\epsilon_{RT})$, and $\chi^2(\epsilon_{RT},\epsilon_{LT})$ for normal ordering. We have checked that the results are similar for inverted ordering as well.
%
We have shown the simultaneous dependence of these GNI parameters using contour plots in Fig.~\ref{fig:TwoParam}. 
The dark orange, light orange, and navy blue regions correspond to $\chi^2$ values within $1\sigma, 2\sigma$, and $3\sigma$, respectively. 

Considering ($\epsilon_{RS},\epsilon_{RT}$) plane from the middle row of Fig.~\ref{fig:TwoParam}, we notice a tiny region around $\epsilon_{RS} = -1, $ and $\epsilon_{RT} = 0.3$ that is disallowed at 1$\sigma$. This can be understood from our one-parameter analysis as shown in Fig.~\ref{fig:OneParam}. There, in the plot of $\epsilon_{RT}$ we notice a peak in $\chi^2$ for $\epsilon_{RT} = 0.
3$ and this creates a synergy in ($\epsilon_{RS},\epsilon_{RT}$) plane and leads to a disallowed region at 1$\sigma$. However, at more than 1$\sigma$, PTOLEMY suffers from continuous degeneracies as can be seen from the contours, which allow all the parameter regions. Similar conclusions can also be drawn from  ($\epsilon_{LS},\epsilon_{LT}$), ($\epsilon_{LS},\epsilon_{RT}$), and ($\epsilon_{RS},\epsilon_{LT}$) planes. From ($\epsilon_{LS},\epsilon_{RS}$) and ($\epsilon_{RT},\epsilon_{LT}$) we notice continuous degeneracies even at 1$\sigma$.
We further notice that the region plots of $\epsilon_{LS}$ vs. $\epsilon_{RS}$, as shown in the topmost panel and region plots of $\epsilon_{RT}$ vs. $\epsilon_{LT}$, as shown in the bottom-right panel have a similar inclination, whereas the other plots do not. This can be understood as a consequence of the interaction terms $\epsilon_{LS}\epsilon_{RS}$ and $\epsilon_{RT}\epsilon_{LT}$ in Eq.~\eqref{eq:BSMamplitude}. It is also to be noted that in Eq.~\eqref{eq:BSMamplitude} there is no interference between $\epsilon_{LS}$-$\epsilon_{LT}$, $\epsilon_{LS}$-$\epsilon_{RT}$, $\epsilon_{RS}$-$\epsilon_{LT}$, and $\epsilon_{RS}$-$\epsilon_{RT}$  and hence these plots do not have any particular orientation.
%

%

%
\section{Electron Spectrum in the Presence of GNI}
\label{sec:results}
\begin{figure}[t!]
\centering
\includegraphics[scale=0.53]{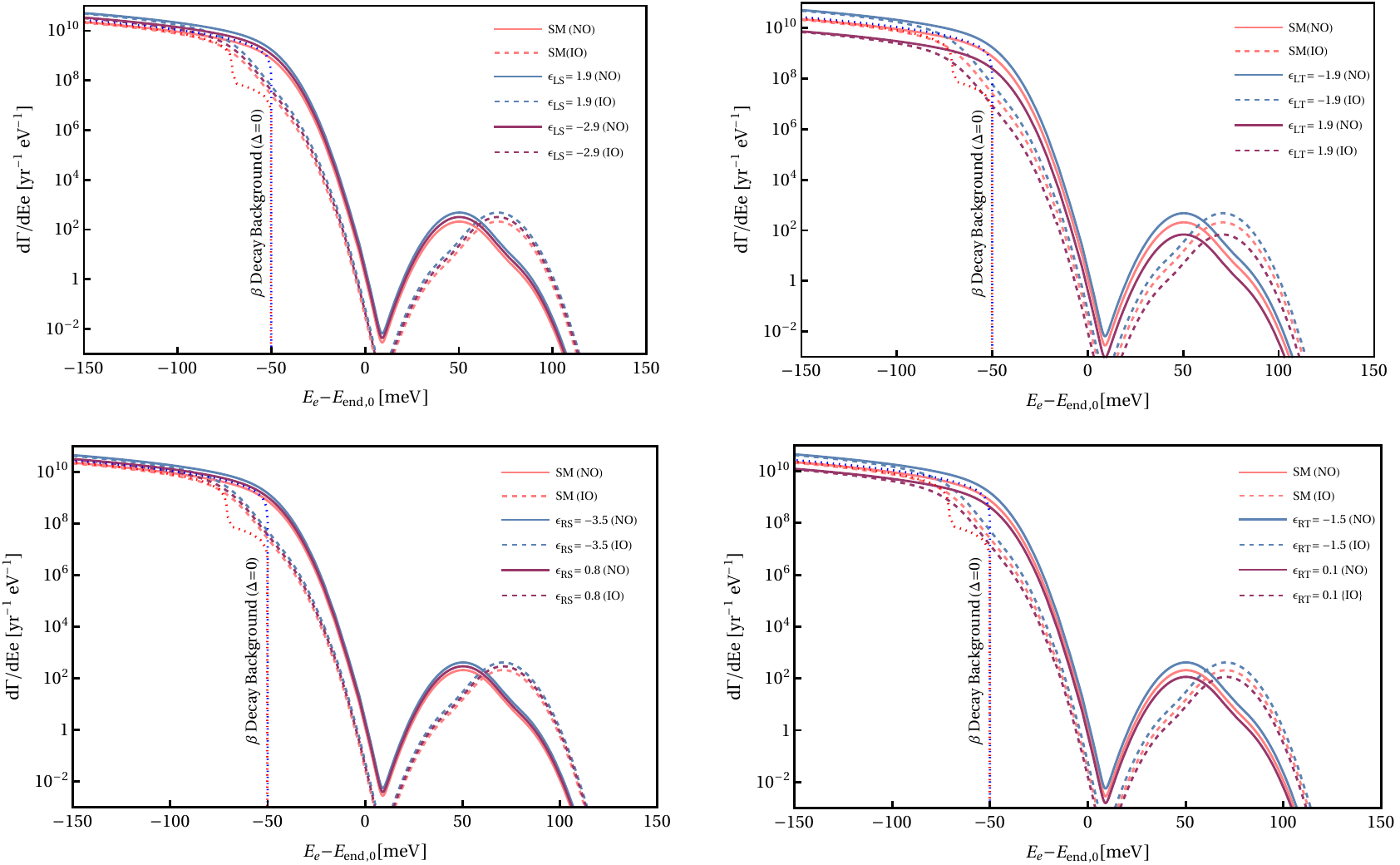}
\caption{\footnotesize The electron spectrum near the end point energy of $\beta$-decay in the presence and absence of GNI for both the mass orderings. The four different plots have four different GNI parameters i.e. $\epsilon_{LS}$, $\epsilon_{LT}$, $\epsilon_{RS}$ and $\epsilon_{RT}$. Here for all four plots $m_{\mathrm{lightest}}=50\mathrm{~meV}$ and $\Delta=20\mathrm{~meV}$. The solid (dashed) lines correspond to the normal (inverted) ordering, whereas the dotted blue (red) line signifies the $\beta$ decay background for the normal (inverted) ordering for $\Delta=0$. We have used the value $\Delta=20\mathrm{~meV}$ as a best case scenario for PTOLEMY. Though this kind of sensitivity is not achievable now, in future this could be the case.}
\label{Fig3}
\end{figure}

In this section, we make a comprehensive analysis to understand  the effect of the GNI parameters on the expected electron spectrum  in CNB measurement using a PTOLEMY detector. The physical observable that is considered in this method is the electron spectrum due to the process 
$
\nu_e + {}^3\mathrm{H} \rightarrow {}^3\mathrm{He} + e^-.
$
It is to be noted that for massive neutrinos, the endpoint energy of the $\beta$-decay background is less than the endpoint energy of massless neutrinos by the mass of the lightest mass eigenstate of the neutrinos, i.e.,
\begin{equation}
E_{\mathrm{end},\mathrm{massive}}=E_{\mathrm{end},0}-m_{\mathrm{lightest}},
\end{equation}
where $E_{\mathrm{end},\mathrm{massive}}$ is the $\beta$-decay endpoint energy for the massive neutrino case.
In Fig.~\ref{Fig3}, we have shown the electron spectrum of neutrino capture on tritium around the endpoint energy of $\beta$-decay for the SM as well as in the presence of the concerned new physics (see figure caption for color details).
%

%
We show the scenario in the absence of finite experimental resolution i.e., for $\Delta = 0$ using the dotted blue and red lines for both the mass orderings. We observe a sharp drop in the $\beta$-decay spectrum from all four panels. For a realistic experimental set-up,  we have smeared the signal with respect to a finite experimental resolution of $\Delta = 20\mathrm{~meV}$, due to which the $\beta$-decay background goes beyond the endpoint energy. For the electron spectrum due to the capture of $\nu_e$, the spectrum peaks beyond the endpoint energy of the $\beta$-decay of tritium.

In order to show the impact of GNI parameters on the electron spectrum of CNB capture in a comprehensive manner, we describe our results in four panels of Fig. \ref{Fig3} considering one non-zero GNI parameter at a time. 
The benchmark values of GNI parameters that are used here are taken from Tab. \ref{table:table4}. The computed electron spectrum corresponds to fixed values of $m_{\text{lightest}}=50\mathrm{~meV}$ and $\Delta=20\mathrm{~meV}$.
We notice from all four panels that the peak of the electron spectrum is at $m_{\mathrm{lightest}}=50\mathrm{~meV}$ for normal neutrino mass ordering (NO), whereas it is at a slightly higher value for inverted ordering (IO).
This can be understood from Eq.~\eqref{eq:CNB-spectrum}. Note that the capture rate peaks at $E_{e}-E_{\mathrm{end},0} =  m_j$. For NO, $j=1$ and hence one gets maximum contribution for $\Gamma_1$ as it depends on  $U_{e1}$ using oscillation data. For IO, $j=3$ we find the least contribution from $\Gamma_3$ using oscillation data on $U_{e3}$.
Besides this, in the second panel of the top row, we have shown the effect of $\epsilon_{LT}$ on the electron spectrum of CNB capture for two benchmark values $\epsilon_{LT} =-1.9,$ and  1.9.
For $\epsilon_{LT} = -1.9$, it can be observed that the number of events is $\sim 2.3$ times higher than the SM scenario, whereas  for $\epsilon_{LT} = 1.9$ the number of events is $\sim 3.3$ times lower than the SM scenario for both NO and IO.
Again, this can be understood from Eq. (\ref{eq:BSMamplitude}). Setting all the GNI parameters to zero, except  $\epsilon_{LT}$, we notice that the larger impact arises for $\epsilon_{LT}=-1.9$   than the SM, whereas the impact is minimum for $\epsilon_{LT}=1.9$.
Similar conclusions can also be made by observing other panels.
%
%
This shows us that the response function of a PTOLEMY-like experiment is equipped to probe the presence or absence of GNIs' effects. It is also interesting that for the values of the GNI parameters which were used in this study, for most cases, the difference between the two scenarios is significant.
We must mention one crucial caveat here that the experimental resolution $\Delta$ must be smaller than the lightest neutrino mass $m_{\text{lightest}}$ to have a distinct signature of the CNB capture from the usual $\beta$-decay background. 
\begin{figure}[t!]
\centering 
\includegraphics[scale=0.53]{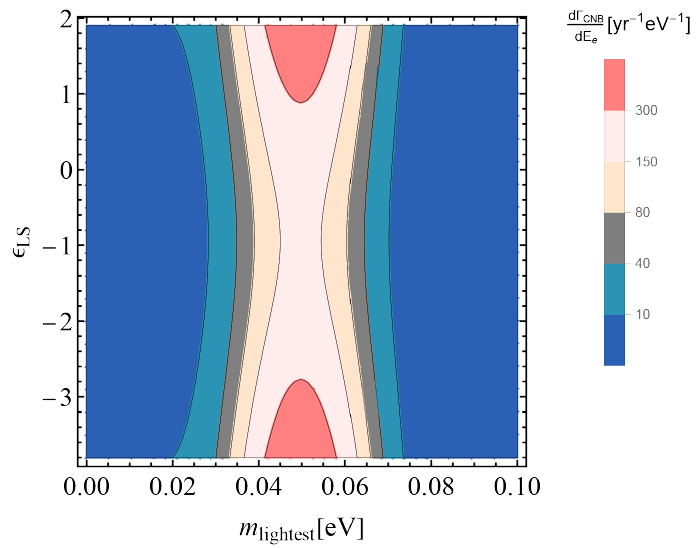}
\includegraphics[scale=0.55]{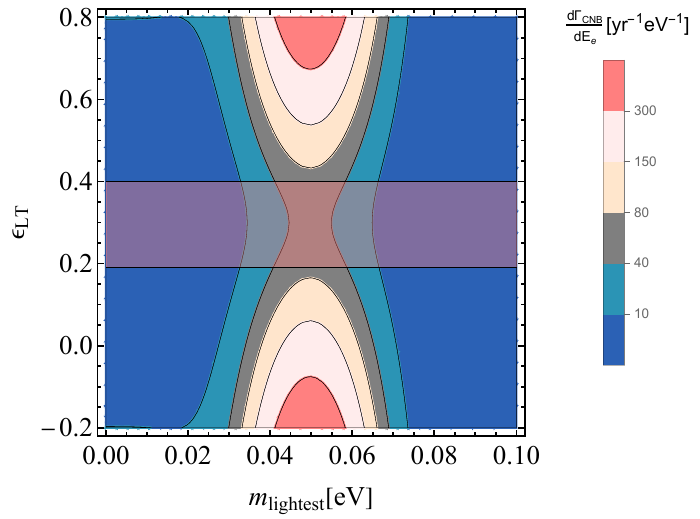}\\
\includegraphics[scale=0.55]{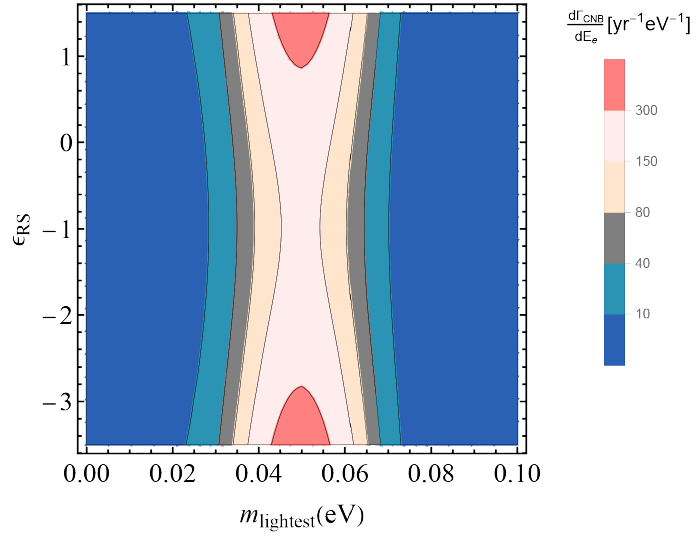}
\includegraphics[scale=0.55]{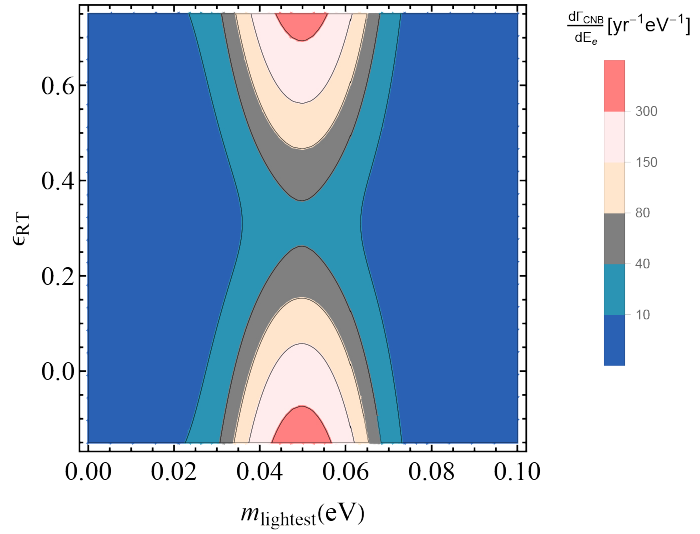}
\caption{\footnotesize Region plots of the expected event rates in the ($m_{\text{lightest}}, \epsilon_{lq}$) plane for normal  ordering  for $E_e-E_{\mathrm{end},0}=50\mathrm{~meV}$.
}
\label{Fig4}
\end{figure}
Unlike Fig.~\ref{Fig3}, where we fixed the values of $m_{\mathrm{lightest}}$ and GNIs, in our following results we have done an analysis to determine the correlation between the $m_{\mathrm{lightest}}$ and the four different GNI parameters by showing region plots of expected event rates. These are shown in Figs.~\ref{Fig4}, and~\ref{Fig5} for NO and IO, respectively. In each panel, the vertical colored bar shows the corresponding rate of CNB neutrino capture.
In all these plots the experimental energy resolution is taken to be $\Delta=20\mathrm{~meV}$.
We notice here that due to the symmetric nature of GNIs (see one parameter  $\chi^2$ analysis as shown in Fig.~\ref{fig:OneParam}), the maximum values of the electron capture rate are observed for two discrete degenerate regions, depicted in red color. These discrete regions can be observed for the expected event rates $ \sim 300$ yr$^{-1}$eV$^{-1}$, whereas the regions become continuous for the number of capture rates below 300 yr$^{-1}$eV$^{-1}$, as shown in the first panel of the top row for $\epsilon_{LS}$. It is apparent from all the panels that the degeneracies of different GNIs depend on the capture rate. For the rate as low as 40 yr$^{-1}$eV$^{-1}$, the discrete degenerate regions have been identified for $\epsilon_{RT}$.
In the top-right panel, we mark the shaded regions using light-red color at 90\% C.L. which corresponds to the  peak in the  $\chi^2$ values as shown in Fig.~\ref{fig:OneParam}. These findings are equally applicable to IO (see Fig.~\ref{Fig5}).
\begin{figure}[t]
\centering 
\includegraphics[scale=0.53]{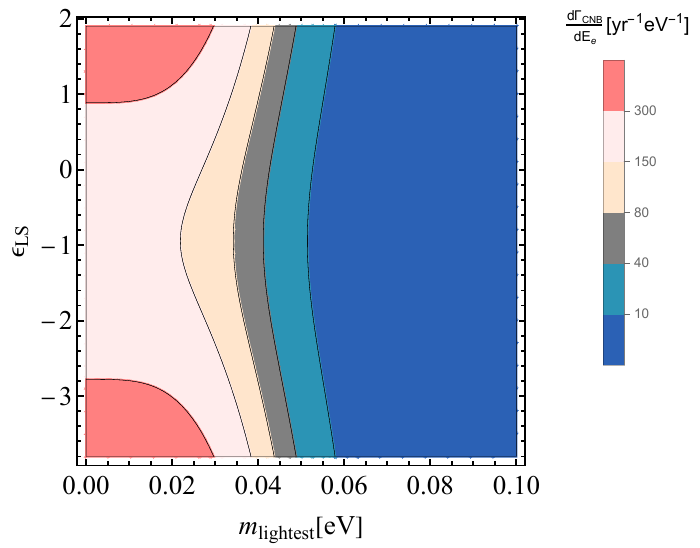}
\includegraphics[scale=0.55]{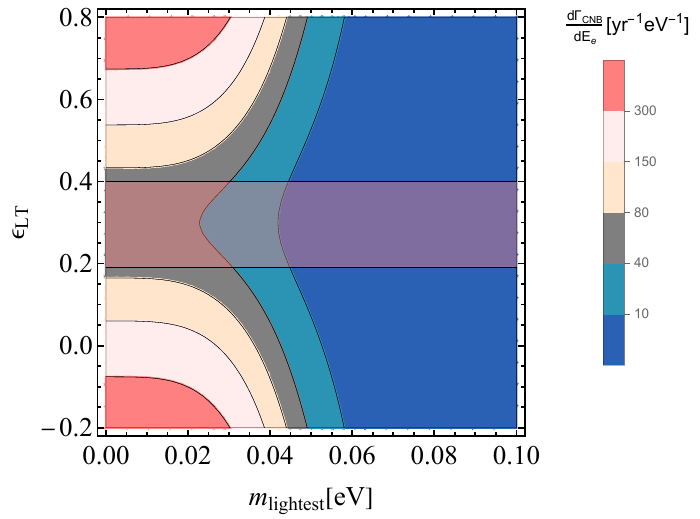}\\
\includegraphics[scale=0.55]{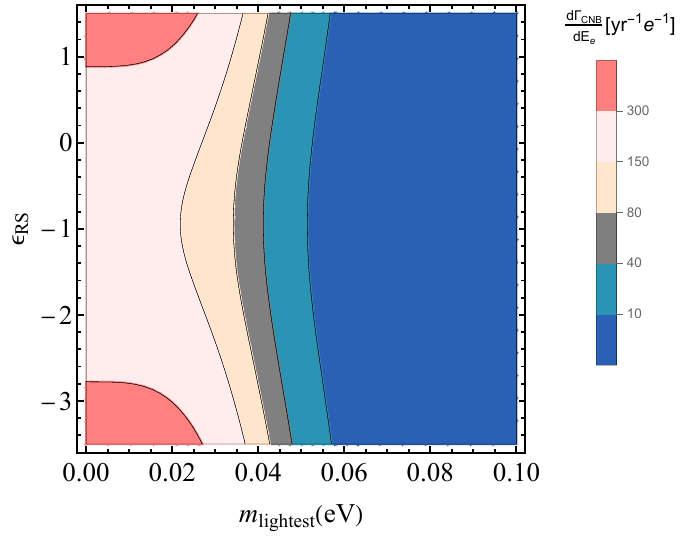}
\includegraphics[scale=0.55]{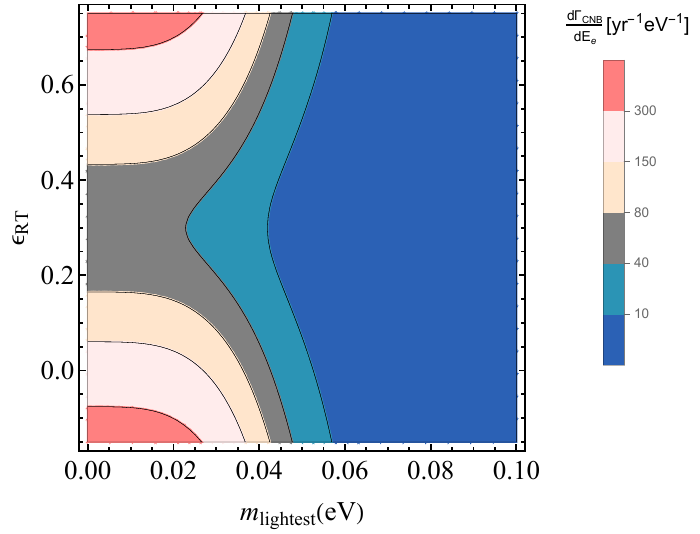}

\caption{\footnotesize Same as Fig.~\ref{Fig4} but for IO.
}
\label{Fig5}
\end{figure}
From Figs.~\ref{Fig4}, and~\ref{Fig5}, a sharp contrast between NO and IO parameter space can be observed. For example, for $E_e-E_{\mathrm{end},0}=50\mathrm{~meV}$, in case of NO maximum event rate can be observed for $m_{\text{lightest}} \sim 50$ meV while in the IO case, it can be observed even for vanishing $m_{\text{lightest}}$.
Besides this, comparing  $\epsilon_{LS}$ plots between NO and IO (see top-left panels of Figs.~\ref{Fig4} and~\ref{Fig5}), it can be observed that for $m_{\mathrm{lightest}}  = 0.03$ eV and  $\epsilon_{LS} \sim \mathcal{O}(0.1)$ the expected rate could be  10 yr$^{-1}$eV$^{-1}$ for NO and for IO, it can reach to 80 yr$^{-1}$eV$^{-1}$. A similar conclusion can also be drawn from the other panels, and hence the PTOLEMY could, in principle, shed light on the ordering of neutrino masses, depending on the allowed parameter space of such exotic couplings. 
%

%
While we are aware that the numerical values for the different couplings in Tab.~\ref{table:table5} are beyond the existing bounds, it is to be noted that we show this to illustrate the capabilities (and limitations) of PTOLEMY. For illustration purposes here we make a few comments about the exact number of CNB events. In the presence of GNI the number of CNB events, i.e. the number of CNB neutrinos captured by PTOLEMY will evidently be different from the SM prediction. For the SM case, for example, if $m_{\mathrm{lightest}}$ is $50\mathrm{~meV}$ and the experimental resolution $\Delta = 20\mathrm{~meV}$ then the number of CNB capture events per year is approximately $4.5$. But this number can vary depending on the presence and the strength of the GNIs considered. If we assume an exposure time of $10$ yr, then for the $\epsilon_{LS}$ case, the lowest observable number of events that will be detected by PTOLEMY is approximately $38$ for $\epsilon_{LS}=-1$. The highest number of events for the $\epsilon_{LS}$ case is approximately $104$ for $\epsilon_{LS}=-3.8$. For different GNIs, the maximum and minimum number of detectable events are mentioned in Tab. \ref{table:table5}.
\begin{table}[t]
\centering
\begin{tabular}{|c|c|c||c|c|}
\hline
\begin{tabular}[c]{@{}c@{}}GNI\\ ($\epsilon_{QX}$)\end{tabular} & \begin{tabular}[c]{@{}c@{}}Minimum no.\\ of evts.\end{tabular} & $\epsilon_{QX}$ (value) & \begin{tabular}[c]{@{}c@{}}Maximum no.\\ of evts.\end{tabular} & $\epsilon_{QX}$ (value) \\ \hline \hline
$\epsilon_{LS}$                                                 & 38                                                             & $-1$            & 104                                                            & $-3.8$          \\ \hline
$\epsilon_{LT}$                                                 & 10                                                             & 0.3             & 108                                                            & 0.8             \\ \hline
$\epsilon_{RS}$                                                 & 37                                                             & $-1$            & 89                                                            & $-3.5$          \\ \hline
$\epsilon_{RT}$                                                 & 7                                                             & 0.31             & 90                                                            & 0.75             \\ \hline
\end{tabular}
\caption{\footnotesize The predicted values of number of CNB events in PTOLEMY for an exposure time of $10$ years. The minimum and the maximum number of events along with the corresponding values of the GNI parameters are mentioned. For an exposure time of 10 years, the SM predicts approximately predicts $45$ CNB events for the same given values, i.e. $m_{\mathrm{lightest}
}=50\mathrm{~meV}$ and $\Delta=20\mathrm{~meV}$.}
\label{table:table5}
\end{table}
%

%
\section{Conclusion}
\label{sec:concl}
In this study, we have explored the realm of GNIs of neutrinos and we have tried to illustrate the impact of them in the detection of CNB through PTOLEMY or a PTOLEMY-like experiment. In this method, we find a prescription to obtain the limitations of PTOLEMY in exploring GNI. After a relook at the usual treatment of CNB detection in the context of SM, we  extend it to accommodate the presence of GNIs.
GNI terms arising due to vector and axial couplings can be absorbed in the CKM elements and hence using only (inverse) $\beta$ decay processes one cannot test these new physics couplings.
Therefore,  we have confined ourselves to the scalar and tensor couplings for both left- and right-handed neutrinos. More precisely, we have considered four GNI parameters, namely, $\epsilon_{LS}$, $\epsilon_{LT}$, $\epsilon_{RS}$ and $\epsilon_{RT}$. 
We perform a $\chi^2$ analysis of the number of events of relic neutrino capture in the presence of GNIs for both one-parameter and two-parameter scenarios. The $90\%$ confidence level values of the four GNI parameters have been taken from the one-parameter $\chi^2$ analysis. For example, $\epsilon_{LS}$ ranges between $[-3.8,1.9]$. Similarly, all the values for the other GNI parameters are mentioned in Tab. \ref{table:table4}. These sets of values are used to obtain the electron spectrum around the endpoint energy of the $\beta$-decay of tritium in the presence of GNIs. These obtained electron spectra, which are shown in Fig.~\ref{Fig3}, give us an idea regarding how the presence of GNI changes the electron spectra. Therefore, analyzing this nature we can get an idea regarding the values of the GNI parameters from the experimental data. We also see that the behavior of the electron spectrum is different for different mass orderings, i.e. the CNB part of the electron spectrum peaks at different energy values for different orderings. More interestingly, this can be used to confirm the neutrino mass orderings or at least put more stringent bounds on the different orderings from the PTOLEMY data. We have also estimated the number of expected events in the presence of GNIs. For an exposure time of $10$ yr, the number of CNB neutrino capture events that can be detected for the SM case is approximately $45$. For the same exposure time, the lowest number of events predicted to be observed is approximately $7$ and that is for $\epsilon_{RT}=0.31$. The highest number of events predicted to be observed is approximately $108$ which corresponds to $\epsilon_{LT}=0.8$.
We have also performed an analysis of the correlation between the mass of the lightest neutrino $m_{\mathrm{lightest}}$ and the GNI parameters $\epsilon_{lq}$ through region plots of $d\Gamma_{\mathrm{CNB}}/dE_e$ for different mass orderings at $E-E_{\mathrm{end},0}=50\mathrm{~meV}$ and the experimental resolution $\Delta=20\mathrm{~meV}$.
It is important to highlight that there are continuous endeavors aimed at attaining an experimental resolution of such magnitude through design modifications. Looking ahead, as more insights into experimental resolution become available, studies of this nature have the potential to deepen our understanding of the lightest neutrino mass and elucidate the ordering of neutrino mass.
This study carried out here can be performed for other neutrino experiments such as KATRIN~\cite{KATRIN:2005fny}, HOLMES~\cite{Nucciotti:2018vyc} etc. In all of these different experiments, there are different features of the relevant neutrinos that will be probed and consequently,  different GNIs can come into the picture. For example, the presence of weak magnetic interaction of neutrinos will be relevant in KATRIN but not play a major role in PTOLEMY. Also, in most of those studies, the non-relativistic approximations have to be discarded in the evaluation of respective cross-sections of these neutrino GNIs. The SM of particle physics needs to be modified to incorporate the answers to all the questions it can not tackle and one of the most efficient ways to do that is by probing these experiments in as many ways as possible. This study could open the doors for many such studies, which not only will give us more insight into new physics but also pave the way for novel probing methods of those physics.

\section*{Acknowledgement}
We thank Ishita Ganguli for collaborating at the initial stage of this work.
The authors are also thankful to M. González-Alonso for his useful suggestions that improved our work significantly.
IKB acknowledges the support by the MHRD, Government of India, under the Prime Minister's Research Fellows (PMRF) Scheme, 2022. NN is supported by the Istituto Nazionale di Fisica Nucleare (INFN) through the ``Theoretical Astroparticle Physics" (TAsP) project. SSS and UKD thanks the erstwhile support from the Department of Science and Technology (DST), Government of India under Grant Reference No. SRG/2020/000283.

 
\bigskip

\bibliographystyle{JHEP}
\bibliography{gniPtolemyRef}
\end{document}